\documentclass{emulateapj}
\usepackage{amsmath}
\usepackage{graphicx}

\shorttitle{Electron Aceleration by the Whistler Instability}
\shortauthors{}

\begin{document}

\title{Stochastic Electron Acceleration by the Whistler Instability in a Growing Magnetic Field}  

\author{Mario Riquelme\altaffilmark{1}, Alvaro Osorio\altaffilmark{1} \& Eliot Quataert\altaffilmark{2}}
\altaffiltext{1}{Departamento de F\'isica, Facultad de Ciencias F\'isicas y Matem\'aticas, Universidad de Chile; mario.riquelme@dfi.uchile.cl}
\altaffiltext{2}{Astronomy Department and Theoretical Astrophysics Center, University of California, Berkeley, CA 94720; eliot@berkeley.edu}

\begin{abstract} 
\noindent  We use 2D particle-in-cell (PIC) simulations to study the effect of the saturated whistler instability on the viscous heating and nonthermal acceleration of electrons in a shearing, collisionless plasma with a growing magnetic field, \textbf{\textit{B}}. In this setup, an electron pressure anisotropy with $p_{\perp,e} > p_{||,e}$ naturally arises due to the adiabatic invariance of the electron magnetic moment ($p_{||,e}$ and $p_{\perp,e}$ are the pressures parallel and perpendicular to \textbf{\textit{B}}). If the anisotropy is large enough, the whistler instability arises, efficiently scattering the electrons and limiting $\Delta p_e$ ($\equiv p_{\perp,e}-p_{||,e}$). In this context, $\Delta p_e$ taps into the plasma velocity shear, producing electron heating by the so called anisotropic viscosity. In our simulations, we permanently drive the growth of $|\textbf{\textit{B}}|$ by externally imposing a plasma shear, allowing us to self-consistently capture the long-term, saturated whistler instability evolution. We find that besides the viscous heating, the scattering by whistler modes can stochastically accelerate electrons to nonthermal energies. This acceleration is most prominent when initially $\beta_e\sim 1$, gradually decreasing its efficiency for larger values of $\beta_e$ ($\equiv 8\pi p_e/|\textbf{\textit{B}}|^2$). If initially $\beta_e \sim 1$, the final electron energy distribution can be approximately described by a thermal component, plus a power-law tail with spectral index $\sim 3.7$. In these cases, the nonthermal tail accounts for $\sim 5\%$ of the electrons, and for $\sim 15\%$ of their kinetic energy. We discuss the implications of our results for electron heating and acceleration in low-collisionality astrophysical environments, such as low-luminosity accretion flows.
\end{abstract} 


\section{Introduction}
\label{sec:intro}

\noindent Nonthermal electrons are usually required to explain observations of various astrophysical systems where MHD turbulence is expected to be present. For example, nonthermal electrons are typically needed to explain the quiescent radio emission in some low luminosity AGNs \citep{LiuEtAl2013} as well as in Sgr A*, the supermassive black hole at the center of the Milky Way \citep{OzelEtAl2000, YuanEtAl2003, BallEtAl2016}. Nonthermal electrons are also necessary to explain the NIR and X-ray emission from Sgr A* flares \citep{YuanEtAl2004,PontiEtAl2017}. In the ICM, the presence of nonthermal electrons is commonly required to explain the extended radio synchrotron emission from some galaxy clusters \citep[see][for a review]{BrunettiEtAl2014}.\newline  

\noindent Several physical processes have been proposed to explain electron acceleration in these systems, including: diffusive shock acceleration \citep[see][for a review]{MarcowithEtAl2016}, magnetic reconnection \citep{SironiEtAl2014,LiEtAl2015}, and various stochastic (second order Fermi) acceleration processes \citep{LynnEtAl2014, ZhdankinEtAl2017}. In general, the disparity between MHD legth-scales and the Larmor radii of particles makes it difficult for MHD turbulence to accelerate electrons via efficient pitch-angle scattering, especially if their energy spectra are initially thermal. Previous works have proposed that this difficulty can be overcome by the resonant scattering provided by whistler waves, producing efficient stochastic electron acceleration in accreting systems \citep{DermerEtAl1996, PetrosianEtAl2004}. In these works, however, the efficiency of the acceleration depends (amongst other parameters) on the spectrum of the whistler fluctuations, which is treated as an input of the models. In this work we use particle-in-cell (PIC) simulations to show that pitch-angle scattering by whistler modes can indeed produce efficient stochastic electron acceleration, in a context in which the whistler waves are consistently generated by magnetic field amplification due to a plasma velocity shear.\newline  

\noindent A key ingredient in our proposed mechanism is electron heating by the so called anisotropic viscosity. This viscous heating arises due to anisotropic pressure tapping into the energy contained in the plasma velocity shear. For an incompressible, homogeneous plasma with no heat flux, the electron internal energy density, $U_e$, changes at a rate \citep{Kulsrud1983, SnyderEtAl1997}:
\begin{equation}
\frac{\partial U_{e}}{\partial t}=q\Delta p_{e},
\label{eq:2}
\end{equation}
where $q$ is the growth rate of the magnetic field ($q=dB/dt/B$, with $B=|\textbf{\textit{B}}|$) and $\Delta p_{e}=p_{e,\perp}-p_{e,||}$ is the difference between the electron pressure perpendicular and parallel to $\textbf{\textit{B}}$. In a collisionless plasma, the difference between $p_{e,\perp}$ and $p_{e,||}$ is a consequence of the adiabatic invariance of the electron magnetic moment, $\mu_e \equiv v_{\perp,e}^2/B$, where $v_{\perp,e}$ is the electron velocity perpendicular to $\textbf{\textit{B}}$. Thus, magnetic field amplification by plasma shear generically drives $p_{\perp,e} > p_{||,e}$, giving rise to electron heating. This process, however, is expected to be limited mainly by the whistler instability \citep{GaryEtAl1996}, which is triggered when $\Delta p_{e}$ surpasses an instability threshold, although ion-scale instabilities can also play a role \citep{RiquelmeEtAl2016}. \newline

\noindent In this work we show that, besides controlling the viscous heating of the electrons, the scattering by the whistler waves can accelerate a fraction of the electrons to energies significantly above thermal. For this, we use 2D particle-in-cell (PIC) simulations of a plasma subject to a permanent shear motion, which continuously amplifies the background magnetic field. Given that this shear motion can be caused by the presence of incompressible MHD turbulence or by differential plasma rotation, our setup mimics a fairly generic physical situation in many turbulent astrophysical systems.\newline  

\noindent In order to optimize our computational resources, we concentrate exclusively on the electron-scale physics. This is done by modeling the ions as infinitely massive particles that only provide a neutralizing electric charge (see \S \ref{sec:numsetup}). This strategy is supported by our previous study of an electron-ion plasma with $2 \lesssim \beta_e \lesssim 20$ ($\beta_e\equiv 8\pi p_e/|\textbf{\textit{B}}|^2$), where the electron anisotropy was mostly regulated by the electron-scale whistler instability, with a moderate contribution from the ion-scale mirror instability \citep{RiquelmeEtAl2016}. For smaller values of $\beta_e$, the effect of the mirror modes is expected to be even smaller \citep{RiquelmeEtAl2015}. As we will see below, this low $\beta_e$ regime is the most interesting in terms of electron acceleration by whistler waves.\newline

\noindent Our paper is organized as follows. \S \ref{sec:numsetup} explains the setup of our simulations. \S \ref{sec:visheating} summarizes the physics of heating by anisotropic viscosity in our simulations. \S \ref{sec:nonthermal} describes the way this viscous heating, through the action of the whistler waves, gives rise to nonthermal electron acceleration. \S \ref{sec:nonthermal} also shows the effects of some plasma parameters on the acceleration efficiency. In \S \ref{sec:conclu} we summarize our results and discuss its astrophysical implications.

\section{Simulation Setup}
 \label{sec:numsetup}
\noindent In this work we use the particle-in-cell (PIC) code TRISTAN-MP \citep{Buneman93, Spitkovsky05} in 2D. In order to make our simulations computationally efficient, we focus exclusively on the electron-scale whistler instability. This is done by using ``infinite mass ions" (the ions are immobile and only provide a neutralizing charge). \newline

\noindent Our simulation boxes consist of a square domain in the $x$-$y$ plane (as shown in Figure \ref{fig:fields}), which contains plasma with a homogeneous initial magnetic field $\textbf{\textit{B}}_0=B_0 \hat{x}$. To amplify the field in an incompressible way, we impose a velocity shear so that the mean particle velocity is $\textbf{\textit{v}} = -sx\hat{y}$, where $x$ is the distance along $\hat{x}$ and $s$ is the shear parameter, which has units of frequency.\footnote{The simulations are performed in the `shearing coordinate system' described in \cite{RiquelmeEtAl2012}, where the shearing velocity of the plasma vanishes, and both Maxwell's equations and the Lorentz force on the particles are modified accordingly.} From flux conservation, the $y$-component of the mean field evolves as $d \langle B_y\rangle /dt = -sB_0$ (throughout this paper, $\langle \rangle$ may represent an average over volume or over particles, depending on the context), implying a net growth of $|\langle \textbf{\textit{B}}\rangle |$. This, due to $\mu_e$ conservation, drives $p_{\perp,e} > p_{||,e}$ during the whole simulation. \newline

\noindent The varying physical parameters between our runs are their initial $\beta_e$ ($\beta_{e,\textrm{init}}$) and the electron magnetization, quantified by the ratio between the initial electron cyclotron frequency and  the shear parameter of the plasma, $\omega_{c,e}/s$.  In our runs we use electron magnetizations that satisfy $\omega_{c,e}/s \gg 1$, but that are still much smaller than expected in real astrophysical settings.\footnote{ As a reference, at $\sim 10$ Schwarzschild radii from Sgr A*, the expected conditions in the accreting plasma imply $\omega_{c,e}/s \sim 10^{11}$ \citep[e. g.,][]{PontiEtAl2017}, where we have approximated $s$ as the Keplerian rotation period at that radius.} Because of this, we made sure to reach the regime where $\omega_{c,e}/s$ is large enough to not to play any role in our final results.\newline
\begin{deluxetable}{llllll} \tablecaption{Parameters of the simulations} \tablehead{ \colhead{Simulation}&\colhead{$\omega_{c,e}/s$}&\colhead{$\beta_{e,\textrm{init}}$}&\colhead{$c/\omega_{p,e}/\Delta_x$}&\colhead{N$_{\textrm{ppc}}$}&\colhead{L/R$_{L,e}^{\textrm{init}}$} } \startdata
  S1 & 1200 &  1 & 15&80 & 110\\
  S2 & 1200 &  2 & 15& 160& 78 \\
  S3 & 3000 &  2 & 15& 160& 78 \\
  S4 & 1200 &  5 & 10& 160 & 74 \\
  S5 & 3000 &  5 & 10& 160 & 74 \\
  S6 & 1200 &  10 & 7&160 & 74  
\enddata \tablecomments{Simulation parameters: the initial electron magnetization $\omega_{c,e}/s$, $\beta_{e,\textrm{init}}$, the electron skin depth $c/\omega_{p,e}/\Delta_x$ ($\Delta_x$ is the grid point separation), the number of electrons per cell N$_{\textrm{ppc}}$, the box size in units of the initial electron Larmor radius $L/R_{L,e}^{\textrm{init}}$. In all runs initially $k_BT_e/m_ec^2=0.28$ and $c=0.225 \Delta_x/\Delta_t$, where $\Delta_t$ is the simulation time step.} \label{table} \end{deluxetable} 
  
\noindent All of our simulations have initially $k_BT_e/m_ec^2 = 0.28$ ($k_B$, $T_e$, and $m_e$ are the Boltzmann's constant, the electron temperature, and the electron mass), which implies $\omega_{c,e}/\omega_{p,e} \approx 0.75/\beta_e^{1/2}$ ($\omega_{p,e}$ is the electron plasma frequency). Their numerical parameters are: the number of macro-electrons per cell (N$_{\textrm{ppc}}$), the electron skin depth in terms of grid point spacing ($c/\omega_{p,e}/\Delta_x$), and the box size in terms of the initial electron Larmor radius ($L/R_{L,e}^{\textrm{init}}$; $R_{L,e}^{\textrm{init}} = v_{th,e}/\omega_{c,e}$, where $v_{th,e}^2=k_BT_e/m_e$). Table \ref{table} shows a summary of our key simulations. We ran a series of simulations ensuring that the numerical parameters do not affect our results. The runs used just for numerical convergence are not in Table \ref{table}.  

\section{The Physics of Electron Heating by Anisotropic Viscosity}
\label{sec:visheating}
 
\begin{figure}[t!]  
\centering 
\includegraphics[width=8cm]{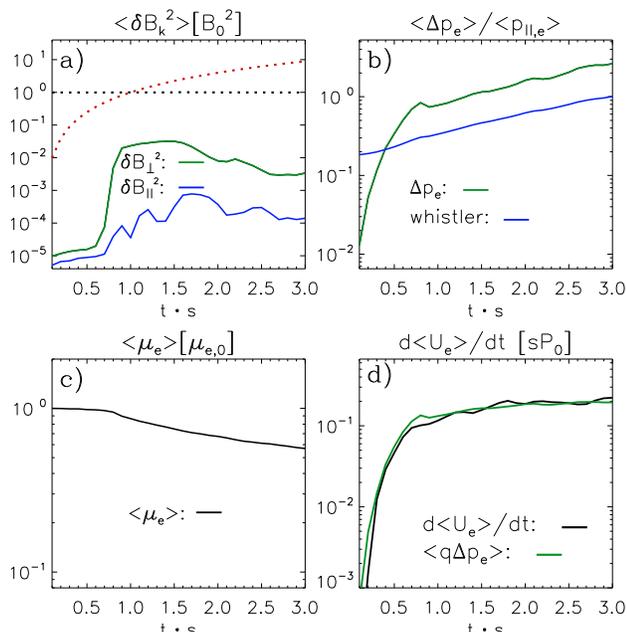}
\caption{The evolution of different volume-averaged quantities for run S2. {\it Panel a:} the energy in $\delta \textbf{\textit{B}}$, normalized by $B_0$ and divided into its components perpendicular ($\delta B_{\perp}$; green) and parallel ($\delta B_{||}$; blue) to $\langle \textbf{\textit{B}}\rangle$. The black-dotted and red-dotted lines show $\langle B_x \rangle^2$ and $\langle B_y \rangle^2$. {\it Panel b:} the electron pressure anisotropy (green line), with the linear whistler instability threshold for growth rates $\gamma_w = 5s$ (blue line). {\it Panel c:} the electron magnetic moment, $\mu_e$. {\it Panel d:} the volume-averaged electron heating rate (black): $d \langle U_e\rangle /dt$, and the expected electron heating rate (green) due to anisotropic viscosity: $\langle q\Delta p_e\rangle$.} 
\label{fig:2d_avquantities} 
\end{figure}

\noindent In this section we describe the process of electron heating by anisotropic viscosity. As a fiducial case we use simulation S2 of Table \ref{table}.
\begin{figure}[t!]  \centering \includegraphics[width=8.6cm]{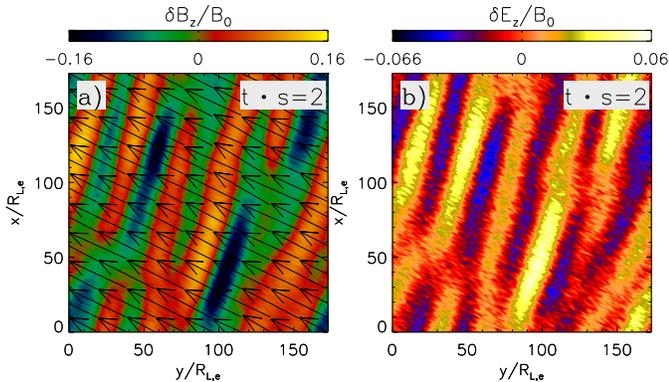}
  \caption{Panels $a$ and $b$ show the $z$-components of the magnetic and electric fields associated with the whistler modes for run S2 at $t\cdot s = 2$. The fields are normalized by the initial magnetic field, $B_0$, and the arrows on panel $a$ show the direction of the magnetic field on the $x$-$y$ plane. It can be seen that the dominant wave vectors of the whistler waves are quasi-parallel to $\langle \textbf{\textit{B}} \rangle$.}
\label{fig:fields}
\end{figure} 
Figure \ref{fig:fields} shows an example of magnetic and electric field fluctuations at $t\cdot s=2$, which naturally arise as we continuously drive the growth of $\langle \textbf{\textit{B}} \rangle$ (we only show the $z$ components). The black arrows in Figure \ref{fig:fields}$a$ represent the projection of $\langle \textbf{\textit{B}} \rangle$ on the $x$-$y$ plane, showing that $|\langle B_y \rangle| \approx 2|\langle B_x \rangle|$, consistent with magnetic flux freezing provided that initially $\langle \textbf{\textit{B}}\rangle=B_0\hat{x}$. Figure \ref{fig:fields} also shows that the field fluctuations are nearly parallel to $\langle \textbf{\textit{B}}\rangle$ and dominated by  wavenumbers $k$ with $kR_{L,e} \sim 0.2$,\footnote{Throughout this paper, $R_{L,e}$ will be a time dependent quantity, which coincides with $R_{L,e}^{\textrm{init}}$ at $t=0$, and takes into account the evolution of $|\langle \textbf{\textit{B}} \rangle|$ and of the rms electron velocity.} both being features that roughly coincide with the characteristics of marginally stable whistler modes \citep{GaryEtAl1996, YoonEtAl2011}.\footnote{The whistler wavenumber at marginal stability satisfies $kR_{L,e}=(\Delta p_e/p_{||,e})^{1/2}(\beta_e/2)^{1/2}$ \citep{YoonEtAl2011}, which means $kR_{L,e}\sim 0.4$ for run S2 at $t\cdot s=2$ (see $\Delta p_e/p_{||,e}$ from Figure \ref{fig:2d_avquantities}).} \newline

\noindent Figure \ref{fig:2d_avquantities} shows how the time evolution of the whistler instability in our shearing setup determines the pressure anisotropy as well as the viscous heating of the electrons. First, Figure \ref{fig:2d_avquantities}$a$ shows how the plasma shear makes $\langle B_y \rangle^2$ (dotted-red) grow while $\langle B_x \rangle^2$ (dotted-black) is kept constant (in our runs $\langle B_z \rangle$ is identically zero). The solid lines show the energy in the fluctuating field $\delta \textbf{\textit{B}}$, separated into its component perpendicular ($\delta B_{\perp}^2$; green) and parallel ($\delta B_{||}^2$; blue) to $\langle \textbf{\textit{B}} \rangle$. We see that there is an initial time (until $t\cdot s\approx 0.7$) where $\delta \textbf{\textit{B}}$ essentially does not change. During this time, $\Delta p_e$ increases linearly, as is shown by the green line in Figure \ref{fig:2d_avquantities}$b$. This evolution of $\Delta p_e$ is caused by the adiabatic invariance of $\mu_e$, as can be seen from Figure \ref{fig:2d_avquantities}$c$.\newline

\noindent This initial regime ends when $\Delta p_e$ reaches the threshold for the growth of the whistler instability. This can be seen by comparing the $\Delta p_e$ evolution with the level necessary for the whistler modes to grow at a rate, $\gamma_{w}$, equal to $5s$ (blue line), obtained from \cite{GaryEtAl1996}.\footnote{We chose the threshold criterion $\gamma_{w}=5s$ since this is the threshold that reproduced best the $\Delta p_e$ evolution in \cite{RiquelmeEtAl2016}, where $ 2 \lesssim \beta_e \lesssim 20$.} After this level is reached, the rapid anisotropy growth stops, and $\Delta p_e$ is maintained in a quasi-stationary state that evolves close to the whistler threshold. The end of the $\mu_e$-conserving regime can also be seen from the exponential growth of $\langle \delta \textbf{\textit{B}}\rangle$ in Figure \ref{fig:2d_avquantities}$a$ at $t\cdot s \sim 0.7$. This growth is dominated by $\delta B_{\perp}$, as expected for the nearly parallel whistler modes. After $t\cdot s\sim 0.7$, the growth of $\langle \delta \textbf{\textit{B}}\rangle$ saturates and its amplitude is maintained at a quasi-stationary level. At $t\cdot s \gtrsim 0.7$, $\mu_e$ decreases at a rate close to $s$ (see Figure \ref{fig:2d_avquantities}$c$), implying that the pitch-angle scattering frequency provided by $\langle \delta \textbf{\textit{B}}\rangle$ is of the order of $s$.\newline

\noindent Although the whistler anisotropy threshold and our obtained $\Delta p_e$ evolve similarly, there is still a factor $\sim 2$ discrepancy. One possible reason for this is the fact that the theoretical threshold is calculated using a bi-Maxwellian electron energy distribution while, as we will see below, in our simulations there is a significant departure from a bi-Maxwellian behavior. Indeed, the presence of a nonthermal, high-energy tail in the electron energy distribution has been shown to make the whistler instability less unstable in the large anisotropy regime relevant for this work \citep{MaceEtAl2010}. Also, 
our mildly relativistic electrons ($k_BT_e/m_ec^2\approx 0.28$) should increase somewhat the pressure anisotropy threshold considered in Figure \ref{fig:2d_avquantities} \citep{BashirEtAl2013}, which is calculated for non-relativistic electrons.\newline

\noindent The electron pressure anisotropy is expected to give rise to viscous heating (see Equation \ref{eq:2}), as has been found by previous PIC simulations \citep[see, e.g., ][]{RiquelmeEtAl2016}. In order to check the importance of this heating mechanism, Figure \ref{fig:2d_avquantities}$d$ shows the volume-averaged electron heating rate (black) for run S2: $d \langle U_e\rangle /dt$, and compares it with the expected electron heating rate due to anisotropic viscosity (green), which is obtained from volume-averaging $q\Delta p_e$. We see that viscous heating accounts for essentially all of the electron heating.

\section{Nonthermal Electron Acceleration}
\label{sec:nonthermal}
\begin{figure}[t!]  
\centering 
\includegraphics[width=9.5cm]{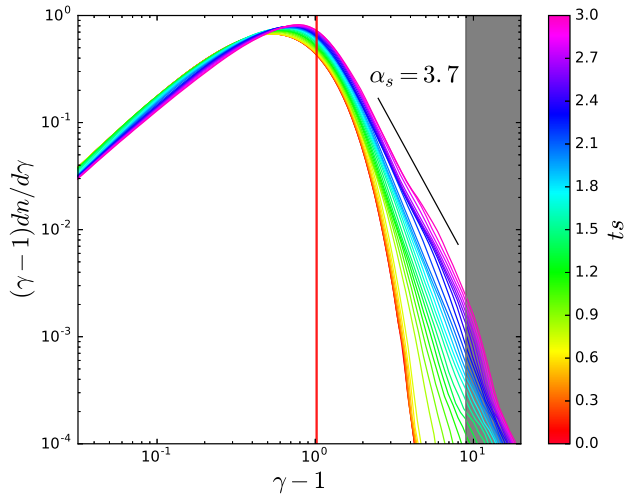}
\caption{The electron energy spectra at different times for S2; the time corresponding to each spectrum is shown by the color bar. After the whistler modes have reached saturation ($t\cdot s \gtrsim 1$), there is a rapid growth of a nonthermal tail. By $t\cdot s=3$ the tail can be modeled by a power-law ($dn/d\gamma \propto (\gamma -1)^{-\alpha_s}$) of spectral index $\alpha_s \sim 3.7$.} 
\label{fig:evol_temp30} 
\end{figure} 

\noindent Figure \ref{fig:evol_temp30} shows the electron energy spectra at different times for run S2, with the time of each spectrum shown by the color bar. After the whistler modes have reached saturation ($t\cdot s \gtrsim 0.7$), there is a rapid growth of a nonthermal tail. By the end of the simulation ($t\cdot s=3$), the tail can be approximated reasonably well  by a power-law of spectral index $\alpha_s \sim 3.7$.\footnote{ This spectral index is estimated using least squares fitting, considering only the range of $\gamma -1$ marked by the black line in Figure \ref{fig:evol_temp30} ($2.5 \lesssim \gamma -1 \lesssim 8$), which gives $\alpha_s=3.69 \pm 0.03$.\label{footnote}}  The magenta lines show that by $t\cdot s=3$ the spectral index is still decreasing, although significantly slower than between $t\cdot s\sim 1$ and $\sim 2.5$, when it experiences its fastest evolution. The peak of the spectrum (representing its `thermal' component) also shifted to larger energies (by a factor $\sim 1.5$) during the simulation, consistent with the overall electron heating.

\subsection{The Acceleration Mechanism}
\begin{figure}[t!]  
\centering 
\includegraphics[width=8.7cm]{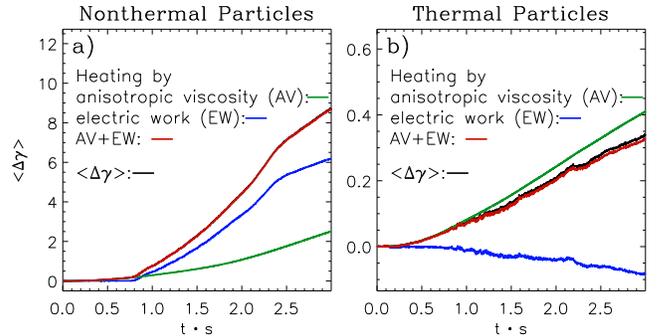}
\caption{Panels $a$ and $b$ show the average change in the electron Lorentz factor $\langle \Delta \gamma \rangle$ (black line) for two groups of `nonthermal' and `thermal' electrons (marked by the grey area and vertical red line in Figure \ref{fig:evol_temp30}, respectively). In both panels, the blue line corresponds to the work done by the electric field associated with the whistler waves, or `electric work' (EW). The green line corresponds to the energy gain by anisotropic viscosity (AV). The red line correponds to the sum of the electric work and the heating by anisotropic viscosity (EW+AV). Notice that in panel $a$ the red and black lines completely overlap.} 
\label{fig:energies} 
\end{figure}

\noindent In order to understand the origin of the nonthermal tail, we calculated the average change in the electron Lorentz factors, $\langle \Delta \gamma \rangle$, as a function of time for two different electron groups. One group represents the electrons from the power-law tail (hereafter, the `nonthermal' electrons), which are picked so that at $t\cdot s=3$ they have $\gamma > 10$. The other group belongs to the bulk of the energy distribution (the `thermal' electrons), and are chosen so that, at $t\cdot s=3$, $0.99 < \gamma-1 < 1$. These nonthermal and thermal electrons are marked by the grey area and by the vertical red line in Figure \ref{fig:evol_temp30}, and their $\langle \Delta \gamma \rangle$ evolutions are shown by black lines in Figures \ref{fig:energies}$a$ and \ref{fig:energies}$b$, respectively. Besides $\langle \Delta \gamma \rangle$, Figures \ref{fig:energies}$a$ and \ref{fig:energies}$b$ also show: 
\begin{enumerate}
\item The work performed by the electric field associated with the whistler waves, hereafter the `electric work' (EW), \footnote{In a shearing plasma there is also an electric field associated to the bulk plasma motion. Since our simulations are performed in the `shearing frame', this electric field vanishes, therefore the electric field in our runs corresponds entirely to the whistler waves.} which is shown by the blue line.
\item The energy gain by anisotropic viscosity (AV), caused by the pressure anisotropy of the electrons,\footnote{For each electron, this energy  gain is calculated by integrating in time the quantity $sv_xv_y\gamma/c^2$, where $v_j$ is the $j$ component of the electron velocity. In a gyrotropic plasma, it can be shown that, when averaged over the electrons, this expression reproduces the heating rate given by Equation \ref{eq:2}.} which is shown by the green line.
\item The sum of the heating by anisotropic viscosity and the electric work (AV+EW), in red line.
\end{enumerate}
\noindent Figure \ref{fig:energies}$a$ shows that the work done by the whistlers' electric field on the nonthermal electrons is more than two times the viscous heating rate. This implies that these electrons are mainly energized through scattering with the whistler waves. On the other hand, Figure \ref{fig:energies}$b$ shows that the energy given by this electric field to the thermal electrons is negative. This implies that the scattering process on average makes the thermal electrons lose energy to the waves. The total heating of the thermal electrons is still positive, and is mainly caused by the dominant (and positive) viscous heating. The reasonably good correspondence between $\langle \Delta \gamma \rangle$ and the sum of the energization by anisotropic viscosity and by the electric work shows that these two processes account fairly well for the total heating of electrons. Also, the fact that the works done by the electric field on the nonthermal and thermal particles are positive and negative, respectively, is consistent with this electric field not affecting the average heating of the electrons, which is dominated by viscous heating (as shown in Figure \ref{fig:2d_avquantities}$d$). \newline 

\noindent In the case of the run S2 analyzed in this section, the nonthermal tail at $t\cdot s=3$ contains $\sim 5\%$ of the electrons and carries $\sim 15\%$ of their energy. Notice that, for the purpose of measuring these numbers, we defined the nonthermal tail through the condition $\gamma -1 > 2$, where the spectrum is fairly well represented by a power-law tail.\newline  

\noindent Thus, in the presented acceleration scenario, the electric field of the whistler waves effectively substracts energy from the bulk of the electron distribution and pass it to a small population of electrons, which get accelerated to nonthermal energies. However, for this mechanism to be efficient, the resonant scattering by whistler modes must occur efficiently for the thermal particles, as well as for the nonthermal electrons in the tail. The resonant condition is
\begin{equation}
\omega - k_{||}v_{||} = \omega_{c,e}/\gamma,
\end{equation}
where $\omega$ is the frequency of the modes, $v_{||}$ is the electron velocity parallel to \textbf{\textit{B}}, and $\omega_{c,e}$ is the non-relativistic electron cyclotron frequency. However, the marginally stable whistler modes satisfy $\omega/\omega_{c,e} =\Delta p_e/(\Delta p_e+p_{||,e})$ \citep{YoonEtAl2011}. Thus, for $\Delta p_e/p_{||,e}\sim 1$ (see Figure \ref{fig:2d_avquantities}$b$), the resonant condition becomes $\omega_{c,e} \sim k_{||}|v_{||}|$. Thus, in order for this condition to be satisfied by the thermal electrons, one needs that $k_{||}R_{L,e}\sim 1$. For the nonthermal electrons a similar requirement needs to be satisfied, but with $k_{||}$ being a factor $\sim 2$ smaller, since the typical value of $|v_{||}|$ for this populations is a factor $\sim 2$ larger than for the thermal electrons (which are the ones that define $R_{L,e}$).\footnote{Indeed, thermal particles in simulation S2 have on average $|v| \sim 0.5c$, while for the nonthermal ones $|v| \sim c$.} This means that, for the nonthermal tail to grow, one requires a moderate widening of the whistler fluctuation power spectrum.\newline 

\begin{figure}[t!]  
\centering 
\includegraphics[width=7.5cm]{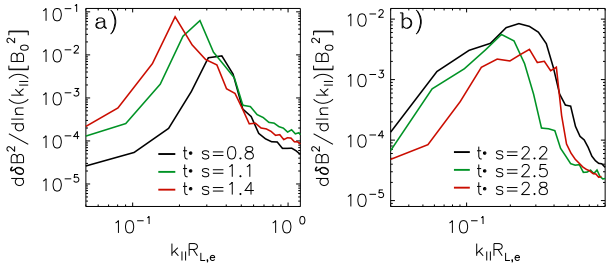}
\caption{The two panels show the power spectra of whistler fluctuations at different times for run S2, expressed as a function of $k_{||}R_{L,e}$. After the saturation of the instability (at $t\cdot s \sim 0.8-1.4$), the spectra are rather narrowly peaked, with their maxima on average at $k_{||}R_{L,e} \sim 0.3$ (see panel $a$).  For $t\cdot s \gtrsim 2$ the spectra widen significantly, having on average significant power between $k_{||}R_{L,e} \sim 0.1$ and 0.4. This allows electrons with a range of energies to resonantly couple to the waves and be stochastically accelerated.} 
\label{fig:spects_timeevol} 
\end{figure} 
\noindent This is indeed what happens in simulation S2, as can be seen from Figure \ref{fig:spects_timeevol}. This figure shows the power spectra of whistler fluctuations at different times ranges: right after the saturation of the instability ($t\cdot s \sim 0.8$; Figure \ref{fig:spects_timeevol}$a$) and after the nonthermal tail is well developed ($t\cdot s \gtrsim 2$; Figure \ref{fig:spects_timeevol}$b$). After saturation, the spectra are rather narrowly peaked, with their maxima on average at $k_{||}R_{L,e}(t) \sim 0.3$.  For $t\cdot s \gtrsim 2$ the spectra widen significantly, having on average significant power between $k_{||}R_{L,e}(t) \sim 0.1$ and 0.4. This result is thus consistent with resonant scattering happening for both thermal and nonthermal electrons. This appears to be a key ingredient to allow the transfer of energy from the thermal to the nonthermal particles. Also, since the whistler electric field is essentially set by the phase velocity of the waves, the proposed acceleration scenario can be interpreted as a stochastic (or second order Fermi) process, given that the scattering with the traveling whistler waves is what ultimately ends up accelerating the nonthermal electrons.\newline 

\noindent In the following sections we analyze how the efficiency of this acceleration depends on different plasma parameters.

\subsection{Dependence on $\beta_{e,\textrm{init}}$} 
\begin{figure}[t!]  
\centering 
\includegraphics[width=8.8cm]{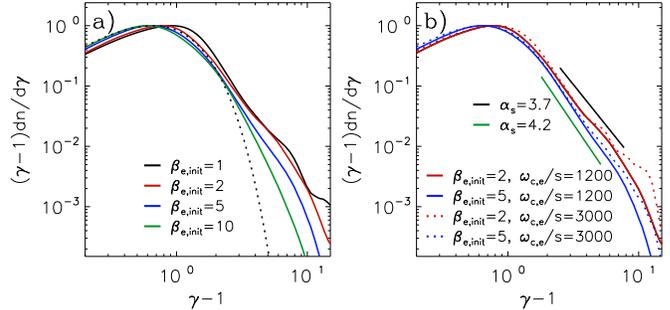}
\caption{{\it Panel a:} the final spectra (at $t \cdot s=3$) of four simulations where the electrons have the same $k_BT_e=0.28m_ec^2$ and $\omega_{c,e}/s=1200$, but $\beta_{e,\textrm{init}}=$ 1 (black), 2 (red), 5 (blue), and 10 (green); the black-dotted line shows the case of a Maxwell-Boltzmann distribution. We see that the non-thermal tails gradually soften as $\beta_{e,\textrm{init}}$ increases; with the $\beta_{e,\textrm{init}}=1$ and 2 cases behaving quite similarly. {\it Panel b:} the spectra for $\beta_{e,init}=2$ and 5 with $\omega_{c,e}/s=1200$ (solid-red and solid-blue, respectively) and with $\omega_{c,e}/s=3000$ (dotted-red and dotted-blue, respectively), along with power-laws of spectral indices $\alpha_s=3.7$ and 4.2 (black and green, respectively; $dn/d\gamma \propto (\gamma -1)^{-\alpha_s}$). The power-laws with indices $\alpha_s=3.7$ and 4.2 provide a reasonably good approximation to the non-thermal tails in the  $\beta_{e,init}=2$ and 5 cases. Also, $\omega_{c,e}/s$ almost does not change the nonthermal tails, implying that the magnetization in our runs does not play a significant role in the electron acceleration.} 
\label{fig:comp_betae} 
\end{figure} 
\noindent Figure \ref{fig:comp_betae}$a$ shows the final spectra (at $t \cdot s=3$) of four simulations where the electrons have the same initial temperature and magnetization ($k_BT_e=0.28m_ec^2$ and $\omega_{c,e}/s=1200$), but different $\beta_{e,\textrm{init}}$. The four cases considered have $\beta_{e,\textrm{init}}=$ 1 (black), 2 (red), 5 (blue), and 10 (green); the black-dotted line shows the case of a Maxwell-Boltzmann distribution. We see that the hardness of the non-thermal tail gradually decreases as $\beta_{e,\textrm{init}}$ increases, with the $\beta_{e,\textrm{init}}=1$ and 2 cases being quite similar. In Figure \ref{fig:comp_betae}$b$ we compare the spectra for $\beta_{e,init}=2$ and 5 (solid-red and solid-blue, respectively) with power-laws of spectral indices $\alpha_s=3.7$ and 4.2 (black and green, respectively; $dn/d\gamma \propto (\gamma -1)^{-\alpha_s}$). We see that these power-law indices provide a reasonably good approximation to the non-thermal tails in these two cases (in the case of $\beta_{e,init}=10$, $\alpha_s \approx 5.5$).\newline

\noindent The softening of the nonthermal tail as $\beta_{e,\textrm{init}}$ increases is in line with the decrease in the overall electron heating, which can be seen from the location of the peak of the spectra in Figure \ref{fig:comp_betae}$a$. Indeed, given that the energy for the formation of the tail is ultimately provided by the viscous heating of the electrons (as we showed in Figure \ref{fig:2d_avquantities}$d$), it makes sense that the nonthermal tail gets less prominent as the viscous heating gets less efficient. This decrease in the viscous heating for larger $\beta_{e,\textrm{init}}$ is expected due to the reduction in $\Delta p_e$ provided by the whistler instability \citep{GaryEtAl1996}.

\subsection{Dependence on $\omega_{c,e}/s$}
\noindent Although in our simulations the electron magnetization satisfies $\omega_{c,e}/s \gg 1$, the values used are still much smaller than the ones expected in realistic astrophysical scenarios. Because of this, we made sure that $\omega_{c,e}/s$ in our simulations is always large enough to not affect our results. We did this for the cases $\beta_{e,init}=2$ and 5, by comparing simulations with $\omega_{c,e}/s = 1200$ and 3000. Our results are shown in Figure \ref{fig:comp_betae}$b$. We see that both for $\beta_{e,init}=2$ and 5 (in red and blue, respectively) there is very little difference between cases with $\omega_{c,e}/s = 1200$ and 3000 (solid and dotted lines, respectively.). The fact that the spectra are essentially independent on $\omega_{c,e}/s$ is consistent with the fact that the time scale for the electron acceleration process is set by the scattering rate by whistler waves. Since in our shearing plasma setup this scattering rate is essentially set by the shear rate $s$ \cite[see][]{RiquelmeEtAl2016}, the expectation is that, at a given value of $t\cdot s$, the electron spectra should be about the same, regardless of the value of $\omega_{c,e}/s$.  

\section{Summary and Discussion}
\label{sec:conclu}
\noindent In this work we used 2D particle-in-cell (PIC) simulations to show that, in the context of a magnetic field amplified by shear plasma motion, the process of electron heating by anisotropic viscosity can give rise to a prominent electron nonthermal tail. Indeed, for $\beta_{e,\textrm{init}}\sim 1$ and after the field is amplified by a factor $\sim 3$,  the electron energy spectrum (measured when $\beta_e\sim 0.1$, because of field amplification) contains a nonthermal power-law tail with spectral index $\sim 3.7$ (see Fig. \ref{fig:evol_temp30}). The nonthermal tail contains $\sim 5\%$ of the electrons, and $\sim 15\%$ of their energy. However, the nonthermal tail becomes progressively softer as $\beta_{e, \textrm{init}}$ grows (see Fig. \ref{fig:comp_betae}). This behavior is in accordance with the decrease in the overall viscous heating of the electrons as $\beta_{e, \textrm{init}}$ increases, since this heating constitutes the ultimate source of energy for the tail formation.\newline

\noindent The role of the whistler instability is key in the electron acceleration process. Firstly, the whistler waves regulate the overall electron energization by determining the efficiency of the electron heating by anisotropic viscosity. This is done by providing an electron pitch-angle scattering rate comparable to the plasma shearing rate, $s$. This keeps the electron pressure anisotropy at a quasi-steady level, which is roughly determined by the whistler instability threshold for growth rate comparable to $s$. Secondly, the whistler waves drive the formation of the nonthermal tail by transferring a significant part of their energy (ultimately gained through the heating by anisotropic viscosity) to a small fraction of electrons. Thus, effectively, the whistler waves subtract energy from the bulk of the distribution, and pass it to the electrons that end up forming the tail. This is inferred from the negative and positive work performed by the whistler electric field on the electrons of the thermal and nonthermal part of the energy spectrum, respectively (see Fig. \ref{fig:energies}). \newline

\noindent The present work focused on mildly relativistic electrons, with initial temperature $k_BT_e=0.28m_ec^2$. Previous theoretical studies have shown that stochastic electron acceleration by whistler waves should be sensitive to the ratio $\omega_{c,e}/\omega_{p,e}$ \citep{DermerEtAl1996, PetrosianEtAl2004}. This sensitivity can be seen as a dependence on $k_BT_e/m_ec^2$, since, for a given value of $\beta_e$, $\omega_{c,e}/\omega_{p,e}$ uniquely fixes $k_BT_e/m_ec^2$.\footnote{\cite{PetrosianEtAl2004} showed that the stochastic acceleration of electrons by whistler waves is more efficient for larger values of $\omega_{c,e}/\omega_{p,e}$, which, for a fixed $\beta_e$, implies larger values of $k_BT_e/m_ec^2$.} We thus defer to a future work the study of the dependence of the presented acceleration process on $k_BT_e/m_ec^2$. It will also be interesting to explore more extensively the regime $\beta_{e,\textrm{init}} \ll 1$, where the dominant whistler modes are expected to adopt a more oblique orientation, which may affect the way the proposed nonthermal acceleration works \citep{GaryEtAl2011}.\newline

\noindent Our proposed acceleration process is likely relevant for low-luminosity accretion flows around black holes, like in Sgr A*. For the specific case of Sgr A*, mildly relativistic electron temperatures and nonthermal spectral components with power-law indices $\alpha_s \sim 3.5$ are favored by multi-wavelength observations \citep{OzelEtAl2000, YuanEtAl2003, BallEtAl2016}. It is important to caution, however, that the proposed factor $\sim 3$ amplification of the local magnetic field may not be so common in the context of a typical turbulent accretion flow. Nevertheless, one could alternatively picture this acceleration to happen due to the cumulative effect of successive (smaller amplitude) growths and reductions of the ambient magnetic field in a turbulent medium. We will study this last possibility in a future work.   

\acknowledgements
\noindent MR and EQ are grateful to the UC Berkeley-Chile Fund for support for collaborative trips that enabled this work. This work was also supported by NSF grants AST 13-33612 and 17-15054, a Simons Investigator Award to EQ from the Simons Foundation and the David and Lucile Packard Foundation. This work used the Extreme Science and Engineering Discovery Environment (XSEDE), which is supported by National Science Foundation grant number ACI-1053575. This research was also partially supported by the supercomputing infrastructure of the NLHPC (ECM-02) at the Center for Mathematical Modeling of University of Chile.

\end{document}